# High-responsivity, High-detectivity Photomultiplication Organic Photodetector Realized by a Metal-Insulator-Semiconductor Tunneling Junction

*Linlin Shi*[1], *Ning Li*[2], *Ting Ji*[1], *Ye Zhang*[1], *Wenyan Wang*[1], *Lin Feng*[1], *Rong Wen*[1], *Yuying Hao*[1], *Kaiying Wang*[1,3], *Furong Zhu*[2], *Guohui Li*[1,\*], *Beng S. Ong*[2,\*], *and Yanxia Cui*[1,\*]

[1]College of Physics and Optoelectronics, Key Lab of Interface Science and Engineering in Advanced Materials, Key Lab of Advanced Transducers and Intelligent Control System of Ministry of Education, Taiyuan University of Technology, Taiyuan 030024, China
[2]Department of Physics, Research Centre of Excellence for Organic Electronics, Institute of Advanced Materials, Hong Kong Baptist University, Kowloon Tong, Hong Kong, SAR, China
[3]Department of Microsystems, University of South-Eastern Norway, Horten, 3184, Norway
E-mail: yanxiacui@gmail.com, liguohui@tyut.edu.cn, bong@hkbu.edu.hk

**Abstract**

Organic photodetectors (OPDs) possess bright prospects in applications of medical imaging and wearable electronics due to the advantages such as low cost, good biocompatibility, and good flexibility. Photomultiplication OPDs (PM-OPDs) enabled by the trap-assisted carrier tunneling injection effect exhibit external quantum efficiencies far greater than unity, thus the acquired responsivities are extremely high. However, the reported PM-OPDs with high responsivity performances are all accompanied by high dark currents due to the introduction of carrier traps, which inevitably results in inferior detectivities. In this work, we modify a P3HT:PCBM donor-rich PM-OPD by introducing an atomically thin $Al_2O_3$ interfacial layer through the ALD technique, obtaining a high responsivity of 8294 A/W and high detectivity of $6.76\times10^{14}$ Jones, simultaneously, both of which are among the highest reported for bulk heterojunction PM-OPDs. Ascribed to the introduction of the atomically thin $Al_2O_3$ layer, the metal-insulator-semiconductor (MIS) tunneling junction is formed, which brings forward a suppressed dark current along with an increased amounts of holes tunneling under forward bias. Meanwhile, the weak light detection limit of the modified PM-OPD within the linear response range reaches the level of $nW/cm^2$. Based on the proposed PM-OPD, a proof-of-concept image sensor with 26×26 pixels is demonstrated, which can respond to both ultraviolet light and visible light. The PM-OPD based sensor arrays can find broad applications for medical imaging, wearable electronics, etc.

Keywords: organic photodetectors, photomultiplication, high-responsivity, high-detectivity, P3HT:PCBM, ALD

**Introduction**

Organic photodetectors (OPDs) have attracted enormous attentions in the areas of medical imaging, wearable electronics, and so on, due to their benefits such as low-cost solution processability, good biocompatibility, good flexibility, light weight, and rich material choices [1-5]. Most of OPDs work on the photodiode mode, with the external quantum efficiencies (EQE) below 100% and the corresponding responsivities relatively low [6-8]. In contrast, the photomultiplication OPDs (PM-OPDs) which possess similar configurations as the photodiode OPDs, can sense light with EQE far exceeding 100%, thereby bringing forward extremely high responsivities In PM-OPDs, the multiplication of electric signals are enabled by the trap-assisted carrier tunneling injection effect. Such a photomultiplication effect is quite attractive in photoplethysmogram, fingerprinting, palmprints, and blood vessel imaging to accurately detect the pianissimo reflected or transmitted light signals from the body, where the light-intensity fluctuation is typically within one-order of magnitude for practical applications[9-11], which can avoid complex amplifier circuits during signal postprocessing owing to its relatively large output electric signal.

However, the realization of high-responsivity in PM-OPDs by introducing more carrier traps, inevitably lead to the increase of current under dark, which deteriorates the device detectivity as well as its weak light detection capability [12-14]. For example, in 2020, a polymer BHJ type PM-OPD made of isolated non-fullerene acceptor was demonstrated with a competitive responsivity of 439.5 A/W but an unimpressive dark current density of 0.05 mA/cm$^2$ under -20 V bias, the combination of which caused an uncompetitive detectivity of $9.5\times10^{12}$ Jones [15]. Similar phenomenon also occurred in PM-OPDs made of polymer donor and fullerene derivative acceptor blends. For instance, a PM-OPD modified on PDPP3T:PC$_{71}$BM displayed a responsivity as high as 761.6 A/W, but its detectivity lies in the level of $10^{12}$ Jones, too, due to the terrible performance under dark [16]. In order to minimize the dark current of PM-OPDs, some endeavors have been made in the past few years [17]. For example, in 2019, the ITO/PEIE substrate with work function of 3.6 eV [18] was applied in PM-OPDs as the cathode, which formed a higher hole barrier with P3HT. Due to the unfavorable hole injection, the dark current of the PM-OPD was apparently decreased. However, the responsivity of that low dark current PM-OPD device was only tens of A/W, much inferior to performances of other reported counterparts [16, 19], implying that the hole injection under illumination was prevented accompanying the minimization of the dark current.

Herein, we propose a universal approach to minimize the dark current while maintaining the responsivity of PM-OPDs, that is by inserting an atomically thin Al$_2$O$_3$ layer between the anode and the hole transport layer. With the help of the Al$_2$O$_3$ interfacial layer between ITO/PEDOT:PSS and donor-rich P3HT:PC$_{71}$BM blend, a series of attractive merits take place due to the formation of a metal-insulator-semiconductor (MIS) tunneling junction. On one hand, the forward biased current under dark can be suppressed by around 6 orders under 8 V bias, facilitating the improvement of detectivity and weak light detection ability. On the other hand, it enables the accumulation of electrons at the side of P3HT:PC$_{71}$BM close to the anode. As a result, more incident photons can be absorbed, with respect to the situation when the electron accumulation takes place at the opposite side, i.e., the side adjacent to cathode. Therefore, the amount of trapped photo-generated electrons can be elevated, bringing forward the injected holes being substantially multiplied and the forward biased responsivity being significantly improved. Overall, the proposed donor-rich P3HT:PC$_{71}$BM PM-OPD possesses an external quantum efficiency, responsivity and detectivity of $3.27\times10^6$ %, 8294 A/W, and $6.76\times10^{14}$ Jones, respectively, all of which are among the highest reported for bulk heterojunction PM-OPDs. Moreover, ascribed to the suppressed dark current, the weak light detection limit of the MIS tunneling junction based PM-OPD reaches the level of nW/cm$^2$, far superior to that of a recently reported PM-OPD (100 nW/cm$^2$) [18]. Based on the proposed PM-OPD with a MIS tunneling junction, a 26×26 pixel image sensor is demonstrated, which can sense light excellently covering both the ultraviolet and the visible wavelength range. It is worth mentioning that, this interfacial engineering method is universal for OPDs based on other polymer blends. The PM-OPD based sensor arrays can find broad applications for medical imaging, wearable electronics, etc. For applications of medical imaging that requires to detect light signals with the power densities greater than $10^5$ nW/cm$^2$ [18], the proposed organic image sensor is quite promising.

**Results and Discussions**

A donor-rich PM-OPD based on the P3HT:PC$_{71}$BM blend (D:A ratio of 100:1) with a simple configuration of ITO/PEDOT:PSS/Blend/Al was studied as the control. This kind of donor-rich PM-OPD was firstly developed in 2015 by Zhang's group[20], which displayed responsivities in the level of hundreds of A/W. Fig. S1(a) and (b) show its schematic diagram and detailed energy levels of different functional layers, respectively. Its current density-voltage (*J-V*) performances under dark and illumination (505 nm, 1.27 mW/cm$^2$) can be found in Fig. 2(a) (squared curves). As displayed in Fig. S1(c), an ohmic-like contact is formed near the anode buffer PEDOT:PSS, and a downward band bending Schottky contact is built up adjacent to the Al cathode[21]. Therefore, the control PM-OPD possesses the unilateral conduction property under dark; see the open squared curve in Fig. 2(a). Holes can be injected from the external circuit into the device under forward bias, but the injection is blocked by a significant amount under reverse bias due to the wide Schottky barrier, as shown in Fig. S1(d) and (e), respectively. Under illumination, the large injection current under forward bias failed to give any distinguishable responses, as can be seen in Fig. 2(a), in agreement with the reported results in literatures[22]. In contrast, under reverse bias, the accumulation of photo generated electrons close to the cathode can narrow the Schottky barrier, so that holes can tunnel into the circuit, thus an EQE greater than 100% can be produced, as illustrated by Fig. S1(f). From the *J-V* curves in Fig. 2(a), it is found that the control device has a light to dark current density ratio ($J_{light}/J_{dark}$) of $3.00\times10^2$, under -8 V. The unimpressive $J_{light}/J_{dark}$ ratio comes from the relatively large dark current. In the control device, both the thermal-activiated-carriers and defect-induced-carriers can flow freely from the blend into the external circuit, i.e., holes across the Ohmic contact and electrons across the Schottky contact, contribute to the poor dark current performance. Due to the same reason, the detectivity along with the weak light detection ability of the control device is limited.

In order to improve the photodetection performance, the PM-OPD device in the configuration of ITO/PEDOT:PSS/Al$_2$O$_3$/Blend/Al is proposed, with the schematic diagram and detailed energy levels as shown in Fig. 1(a) and Fig. S2, respectively. Here, an atomically thin Al$_2$O$_3$ insulator layer of 0.8 nm thick is intentionally introduced in-between the anode buffer and the donor-rich P3HT:PC$_{71}$BM blend by the Atomic Layer Deposition (ALD) technique[23]. Therefore, on the anode side of the device, a metal-insulator-semiconductor (MIS) tunneling junction is formed[24]. In Fig. 2, the dotted curves represent the photodetection performances of the proposed PM-OPD with Al$_2$O$_3$. It is seen clearly from Fig. 2(a) that, the dark current density of the MIS tunneling junction based PM-OPD (shortened as MIS based PM-OPD) device is reduced significantly when either forward or reverse biased. Particularly, the dark current has been lowered by around six orders and one order under 8 V bias and -8 V bias, respectively. As witnessed from the photo current density curves measured under illumination, the existence of the MIS junction only weakly reduces the photo current under reverse bias, and the photo current response under forward bias is higher than that under reverse bias. The $J_{light}/J_{dark}$ ratio of the proposed MIS based PM-OPD reaches $1.56\times10^5$ under 8 V bias, about 500 times superior to that of the control device measured under -8 V bias ($J_{light}/J_{dark} = 3.00\times10^2$).

With the help of the MIS tunneling junction, the EQE and responsivity are improved over a broadband wavelength range from 320 nm to 680 nm, as displayed in Fig. 2(b) and (c). In the spectral experiment, the monochromatic light is supplied by a Xe lamp with the power densities at varied wavelengths shown in Fig. S3 and the applied bias is 19 V. It is found that, the EQE of the proposed PM-OPD at the wavelength of 315 nm reaches $3.27\times10^6$ %, corresponding to a responsivity of 8294 A/W, which are more than 48 times higher than those of the control (EQE: $6.71\times10^4$ %, responsivity: 170.1 A/W, bias: -19 V at 325 nm). The achieved performances are superior to all reported values of PM-OPDs in literatures, as listed in Table 1. Indeed, the outstanding responsivity in the ultraviolet wavelength range are more than four orders higher than those of commercial GaN or SiC photodetectors (0.2 A/W) [25, 26]. Moreover, the synergistic effect of the photo current elevation and dark current reduction is expected to yield an improved detectivity, which is the figure of merit characterizing the capability of a photodetector to detect weak illuminations. Fig. 2(d) shows the measured noise current, the derived shot noise and the thermal noise of the proposed PM-OPD device (see the methods in Supporting Information), which indicates that the total noise is dominated by the shot noise. Thereby, the device detectivity can be approximately determined from the responsivity and the dark current; see in Fig. 2(c). It is witnessed that, besides of an increased responsivity, the MIS based PM-OPD possesses a significant increase of detectivity over the broadband range; and the detectivity reaches $6.76\times10^{14}$ Jones at 315 nm, among the highest value of all PM-

OPD devices; see in Table 1.

To understand the mechanism of the dark current inhibition, the analysis on the energy band diagrams of the proposed MIS based PM-OPD when unbiased or biased are carried out, as displayed in Figs. 1(b)-(f). As shown, in addition to the downward band bending Schottky junction formed close to the Al cathode, the MIS tunneling junction is built up on the anode side. The downward band bending caused by the depletion of holes at the $Al_2O_3$/P3HT:PCBM interface fulfills the prerequisite of rejecting the holes from the external circuit into the device under forward bias, as shown in Fig. 1(e), similar to the effect of the downward band bending Schottky junction at the P3HT:PCBM/Al interface under reverse bias (see in Fig. 1 (c)). This depletion region can only be realized by an insulator that is continuous and dense. Fig. S3 shows the dark current performances of the MIS based PM-OPDs with different $Al_2O_3$ thicknesses. As can be seen in Fig. S4, a 0.2 nm thick $Al_2O_3$ cannot form a continuous insulating film so that the PM-OPD still conduct unilaterally. With the increase of the $Al_2O_3$ layer thickness, the dark current first decreases and then increases. In principle, the thicker $Al_2O_3$ insulator would inhibit the tunneling of carriers through it, leading to a reduction of the dark current. However, an over thicker $Al_2O_3$ film would withstand a higher voltage, resulting in the decrease of barrier height of the downward band bending at the $Al_2O_3$/P3HT:PCBM interface [23, 24, 27]. The lower barrier height could weaken the carrier blocking effect in the depletion region close to $Al_2O_3$, causing the increase of the dark current. With the competition of the two effects, an optimal dark current performance of the proposed PM-OPD is produced when the $Al_2O_3$ thickness is 0.8 nm. We also measured the frequency dependent impedances of the proposed device and the control under -8 V bias. As can be seen in Fig. 2(e), an additional MIS tunneling junction provides a greatly increased shunt resistance, indicating that the leakage current arising from the carrier flow from the blend into the anode has been successfully suppressed.

For comparison, the PM-OPDs with a polymethyl methacrylate (PMMA) insulating film inserted at the PEDOT:PSS/P3HT:PCBM interface were also fabricated with the schematic diagram and detailed energy levels as shown in Fig. S5(a-b). Unfortunately, this PMMA based device cannot suppress the dark current under forward bias and it conducts unilaterally; see in Fig. S5(c). As known, the PMMA film is an inter-cross-linked network film [28-30], so that it is not enough dense to produce the carrier accumulation or depletion phenomenon, thus the initial ohmic-like contact is maintained near the anode buffer PEDOT:PSS. Such a phenomenon deviates from the function of the $Al_2O_3$ film. In other words, the failure on forming an effective MIS tunneling junction causes the holes be injected from the ITO electrode into the internal circuit by a large amount, as shown in Fig. S5(d). From Fig. S5(c), it is seen that, introducing an additional 0.4 nm $Al_2O_3$ on top of the PMMA film can recover the suppression of the dark current under forward bias, confirming that an atomically thin $Al_2O_3$ insulating film that is continuous and dense is essential to form the required MIS tunneling junction.

Next, the working mechanisms of the proposed MIS based PM-OPD under illumination are analyzed, with its energy band diagrams being displayed in Figs. 1(d-f), respectively. Under reverse bias, the accumulation of photo generated electrons close to the cathode can narrow the Schottky barrier, so that the hole injection from the Al cathode into the proposed PM-OPD is realized, similar to that in the control device. On the anode side, the MIS junction barrier may inhibit the holes flowing through to some extent. However, the atomically thin $Al_2O_3$ film allows the carriers to directly tunnel through; meanwhile, the depletion region of the MIS junction can also be narrowed due to the squeeze of the photo generated holes towards the $Al_2O_3$/P3HT:PCBM interface. These behaviors result in the reduction in photo current much less severe than that that in dark current. Similar to the situation of reverse bias, the downward band bending in the MIS tunneling junction functions as a switch of hole tunneling under forward bias. In detail, when forward biased, it is the narrow of the MIS tunneling junction barrier induced by the accumulation of photo generated electrons enables the hole tunneling from the external circuit into the device, so that the photomultiplication response with EQE much greater than 100% can be produced. Again, the photo generated holes are expected to squeeze towards the Al cathode so that the depletion region of the Schottky junction gets shrunk under illumination. Therefore, the current flow arising from the hole injection from the anode that ensures the giant photomultiplication effect can flow smoothly from the blend towards the cathode. [31, 32]

To understand the reason why the photo current of the MIS based PM-OPD under forward bias is greater than the value measured when the same voltage is applied oppositely, the optical field simulations are carried out. As known, the accumulation of photo generated electrons close to the carrier tunneling electrode is the prerequisite for the following hole tunneling process, therefore only the photo absorption close to that electrode can contribute to the following multiplication effect. Fig. 2(f) shows the distribution of the photo generated carriers versus wavelength for the proposed PM-OPD, and it is observed that the absorption of light in the P3HT:PCBM region close to the anode is much stronger than that close to the anode, perfectly explaining why the photo current response under forward bias is superior to that under reverse bias. Overall, the MIS tunneling junction not only plays a role on suppressing the dark current, but also makes use of the photo generated carriers as much as possible. It is the combination of the two beneficial effects brings forward the attractive photodetection performances of the MIS based PM-OPD.

Since the dark current of the PM-OPD has been reduced significantly by the incorporation of the MIS tunneling junction, it is expected that both the LDR and the weak light detection ability are also improved. LDR is derived from the difference of the highest and the lowest detectable illumination power densities ($P_{high}$ - $P_{low}$) within the linear response range. Figs. 3(a-b) show the photocurrent density versus illumination power intensity for the control device under -8 V, and the MIS based device under a bias of -8 V and 8 V, respectively. The measurements were carried out when the devices were under illumination of a 532 nm continuous laser with different power densities ($P_{in}$). As can be seen, by inserting the $Al_2O_3$ interfacial layer, the LDR of the

PM-OPD is broadened from 93 dB to around 110 dB. The transient responses of different PM-OPDs under illumination of 640 nW/cm$^2$ and 64 nW/cm$^2$ are displayed in Figs. 3(d-e). It is found that the proposed PM-OPD with the MIS tunneling junction can respond stably with strong signal contrast when the illuminations are turned on and off. Instead, the control device exhibit a very tiny difference between the current signals under illumination and dark when the illumination power density is 640 nW/cm$^2$. The control device completely loses the signal contrast when the illumination power density is 64 nW/cm$^2$, under -8 V bias.

It is noticed that the LDR of the MIS based PM-OPD under 8 V is a bit smaller than that under -8 V, and there is an obvious turning point ($P_{turning}$ = 64 nW/cm$^2$, equal to $P_{low}$) in the LDR plot under 8 V. Below the turning point, although the proposed PM-OPD can still sense light with apparent signal contrasts, the curve of the photocurrent density versus illumination power intensity deviates from the linear response range. If the applied voltage increases to 19 V, the turning point $P_{turning}$ can be reduced by more than one order, from 64 nW/cm$^2$ to 2.5 nW/cm$^2$, as shown in Fig. 3(c). As a result, the LDR of the MIS based PM-OPD can be broadened to 116 dB. The transient response of the MIS based PM-OPD under illumination of 2.5 nW/cm$^2$, when the applied bias is 19 V, is displayed in Fig. 3(f). As can be seen, the proposed PM-OPD responds stably with a strong signal contrast when the weak illumination ($P_{in}$: 2.5 nW/cm$^2$) is turned on and off. For comparison, the response of the control device under such a weak illumination, when the applied bias is -19 V, is also presented in Fig. 3(f). As seen, unsurprisingly, the control device can only output a straight current line no matter how the illumination is changed. According to these observations, it is concluded that due to the suppressed dark current, the weak light detection ability of the MIS based PM-OPD is improved obviously with respect to the control. The acquired weak light detection performances ($P_{low}$:2.5 nW/cm$^2$, LDR: 116 dB) in our work are much superior to those of the recently reported PEIE based PM-OPD ($P_{low}$:100 nW/cm$^2$, LDR: 60 dB) [18], and they are comparable to those inorganic photodetectors, such as GaN and InGaAs [33, 34].

The existence of the turning point in the LDR plot of the MIS based PM-OPD device implies that the current injection behavior experiences a transformation when the illumination power density enters the LDR range. It is known from Fig. 1(f) that the narrow of the MIS tunneling junction barrier induced by the accumulation of photo generated electrons enables the hole tunneling from the external circuit into the device. The fact is that, only the photo generated electrons with a sufficiently rich amount can induce a MIS junction barrier with an enough narrow width ($W_T$), as indicated in Fig. 5(a), so that the injected holes can directly tunnel through the barrier with negligible loss. If the power density of the illumination is low, the width of induced MIS junction barrier width ($W$) will be greater than $W_T$, as shown in Fig. 5(b). Then, the flow of holes across the barrier would suffer a significant amount of loss, resulting in the curve of the current density versus the illumination power density deviates from the linear rule. It is reflected from Figs. 3(b) and (c) that the turning point in the LDR plot of the MIS based PM-OPD device depends on the amplitude of the applied forward bias, and the greater the forward bias, the lower the amplitude of $P_{turning}$. This is because, with the increase of the applied bias, the electron accumulation becomes easier to be completed ascribed to the improved electron drift property. It is emphasized that because the dark current has been suppressed to a very low level, the PM-OPD can sense light that is weaker than $P_{turning}$. For example, the MIS based PM-OPD can rise and fall following the illumination on and off signals when the power density is 6.4 nW/cm$^2$ and 1.2 nW/cm$^2$, under 8 V bias and 19 V bias, respectively, as displayed in Figs. 4(c-d). As observed, the weak light signals induce current responses very slowly, because the carrier accumulation process takes time to complete. This is distinct from the response under illumination with high power densities, for which, the electron accumulation can be finished very fast. Fig. S6 shows the transient response of the MIS based PM-OPDs under illumination of 1.27 mW/cm$^2$, from which the response time is deduced to be 83.59 ms, superior to the results of other PM-OPDs made of the donor-rich P3HT:PCBM blend [18, 35]. On the basis of the individual PM-OPD, we develop a 26×26 pixel organic image sensor with the diagram representing the setup of the scanning process shown in Fig.5(a). A certain pixel is selected when its bottom anode line and top cathode line are applied with a bias, and then the photocurrent is readout. The schematic diagram of the crossed electrode lines, and the physical map of the top cathode lines are shown in Figs. 5(b-c), respectively. Fig. 5(d) displays a part of the optical image of the 26×26 pixel organic image sensor. As can be seen, the active area of the overlap between the ITO anode and Al cathode is 200 μm×200 μm. Later, A "T" shape shadow mask was placed in contact with our image sensor for verifying the imaging function. To reconstruct the optical pattern projected on the image sensor, a photocurrent value was converted to grayscale number between 0 and 255, with the conversion formula being given in Supporting Information. Fig. 5(e) show the images of the object under illumination of 375 nm, 505 nm, 565 nm, and 660nm at a power density of 1.27 mW/cm$^2$. It can be seen that the symbol "T" can be clearly resolved, indicating that the imaging function of the proposed broadband organic sensor device with the responding wavelength range covering from ultraviolet to visible is reliable.

At last, the proposed method of forming a MIS tunneling junction with the help of an atomically thin $Al_2O_3$ interfacial layer is universal for improving the performances of other polymer OPD devices. An instance was demonstrated for the device in the configuration of ITO/PEDOT:PSS/Blend/Al with the blend made of PTB7:PCBM (100:1). It is mentioned that the PTB7:PCBM (100:1) blend cannot induce the photo multiplication effect. As can be seen in Fig. S7, for the control device, the photo current and the dark current almost coincide. As a contrast, by introducing a 0.8 nm thick $Al_2O_3$ interfacial layer, the photo current under 505 nm LED at 1.27 mW/cm$^2$ can be apparently distinct from the dark current under both forward and reverse bias. This confirms that the thin $Al_2O_3$ interfacial layer can facilitate the formation of the MIS tunneling junction, thereby suppressing the dark current by a large extent in OPDs.

**Conclusions**

In summary, we have reported a high-responsivity, high-detectivity PM-OPD by incorporating an atomically thin $Al_2O_3$ interfacial layer between the hole transport layer

PEDOT:PSS and the donor-rich blend. Ascribed to the introduction of the atomically thin $Al_2O_3$ layer through the ALD technique, the MIS tunneling junction has been built up. As result, a suppressed dark current along with an increased amount of holes tunneling under forward bias have been simultaneously achieved. It has been demonstrated the proposed PM-OPD exhibited an EQE, responsivity, and detectivity of $3.27 \times 10^6$ %, 8294 A/W, and $6.76 \times 10^{14}$ Jones, respectively, which are all among the highest of reported PM-OPDs. The incorporation of a MIS tunneling junction also improves the ability of weak light detection. The transient current response measurement implied clearly that the MIS tunneling junction based PM-OPD could sense light as weak as a few $nW/cm^2$; in contrast, the control device could only respond to illumination with the power densities two order higher. We also fabricated a proof-of-concept PM-OPD image sensor with 26×26 pixels, which could reliably reconstruct the input symbol over a wavelength range from the ultraviolet to visible. The results in this work are encouraging by providing in-depth insights into the design and implementation of high-responsivity and high-detectivity OPDs for applications of medical imaging, wearable electronics, and so on.

**Acknowledgements**

The authors are grateful for support from the National Natural Science Foundation of China (No. 61922060，61775156，61805172，and 61905173), the Graduate Innovation Project of Shanxi Province (No. 2020BY117), the Key Research and Development (International Cooperation) Program of Shanxi Province (No. 201803D421044), the Natural Science Foundation of Shanxi Province (No. 201901D211115), the Henry Fok Education Foundation Young Teachers Fund, Transformation Cultivation Project of University Scientific and Technological Achievements of Shanxi Province (No. 2020CG013), Introduction of Talents Special Project of Lvliang City, and Platform and Base Special Project of Shanxi Province (No. 201805D131012-3).

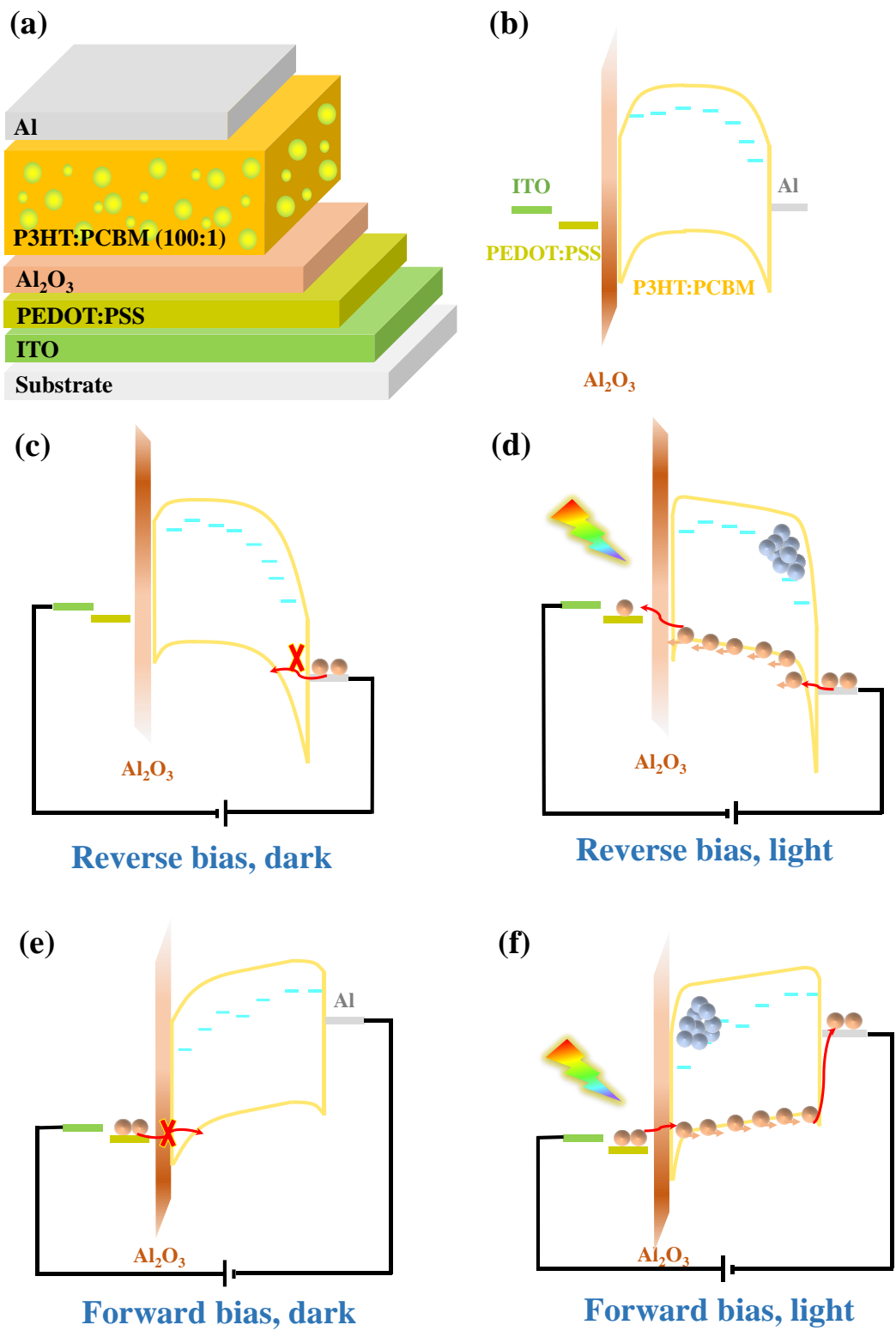

Fig. 1 (a) Schematic diagram of the MIS junction based PM-OPD formed by introducing an atomically thin $Al_2O_3$ layer. (b-f) Energy band diagrams of the MIS junction based PM-OPD. The device is unbiased (b), reverse biased (c-d), and forward biased (e-f). In (c) and (e), the device is in dark, and in (d) and (f), it is under illumination.

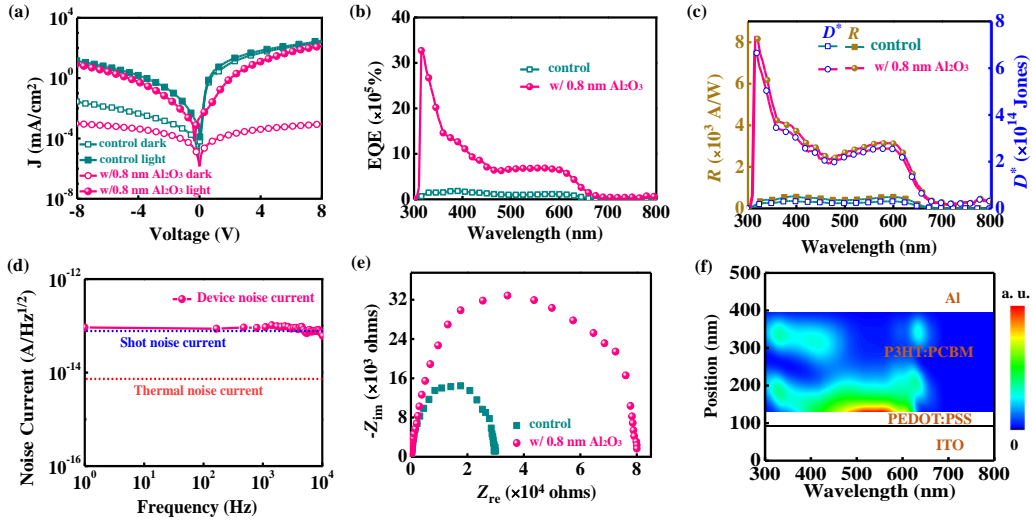

Fig. 2 (a) *J–V* curves of the control and the MIS based PM-OPD devices in dark and under 505 nm LED illumination at a power density of 1.27 mW/cm$^2$. (b-c) External quantum efficiency, responsivity, and detectivity of the control (-19 V bias) and MIS based PM-OPD (19 V bias). (d) Frequency-dependent noise current for the MIS based PM-OPD. (f) Electrochemical impedance spectroscopy (EIS) curves of the control and the MIS based PM-OPD. (f) Wavelength-dependent absorption distribution map in the PM-OPDs.

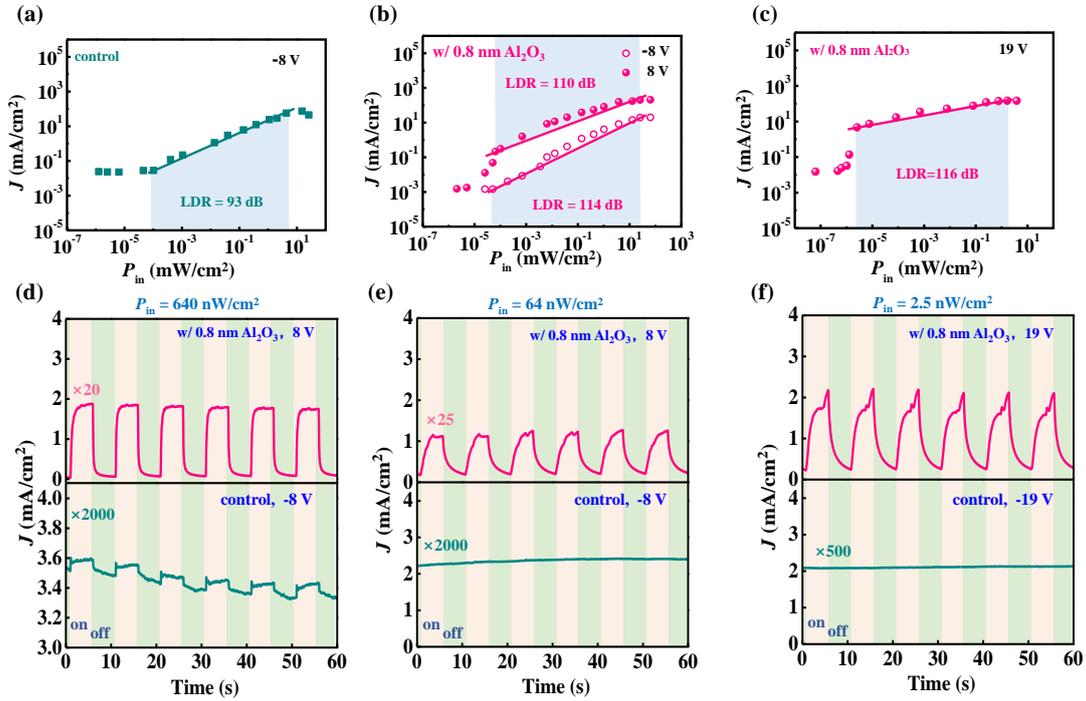

Fig. 3 (a-c) LDR plots of the control and MIS based PM-OPDs. (d-f) Transient responses of the control and MIS based PM-OPD at different illumination power intensities (640 nW/cm$^2$, 64 nW/cm$^2$, and 2.5 nW/cm$^2$). The reverse bias amplitude for the control and the forward bias amplitude for the MIS based PM-OPD are 8 V in (a-b) and (d-e), and they are 19 V in (c) and (f).

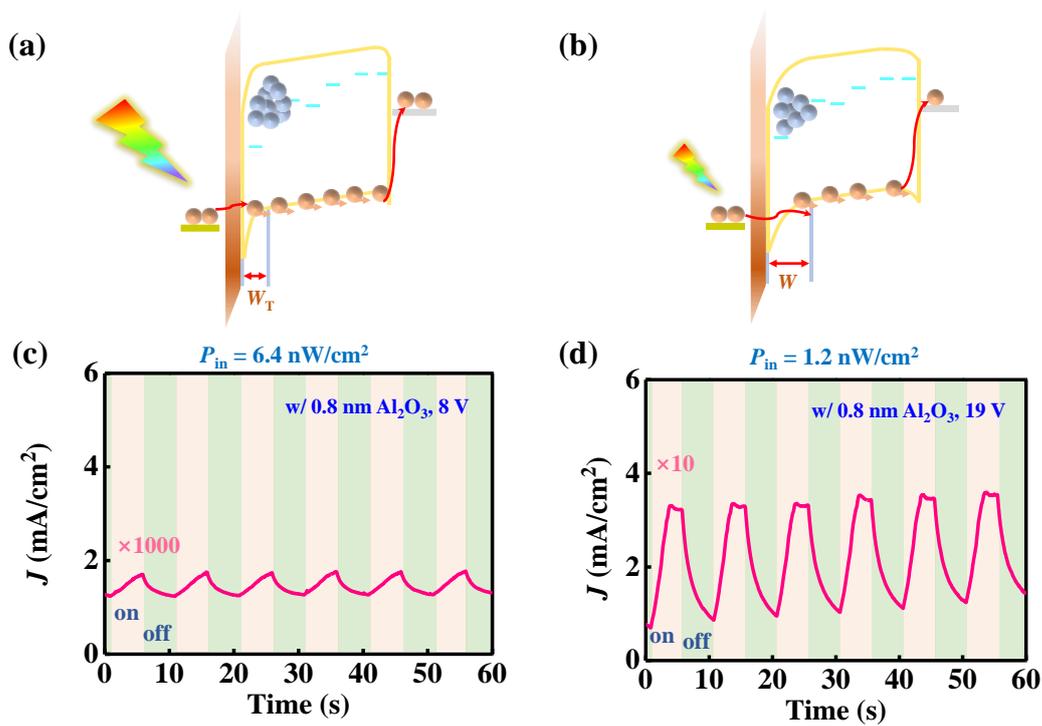

Fig. 4 Energy band diagrams of the MIS based PM-OPD when the illumination locates at the turning point ($P_{turning}$) of LDR (a) and is lower than $P_{turning}$ (b), respectively. Transient responses of the MIS based PM-OPD with the illumination power intensity of 6.4 nW/cm² under 8 V (c) and 1.2 nW/cm² under 19 V (d).

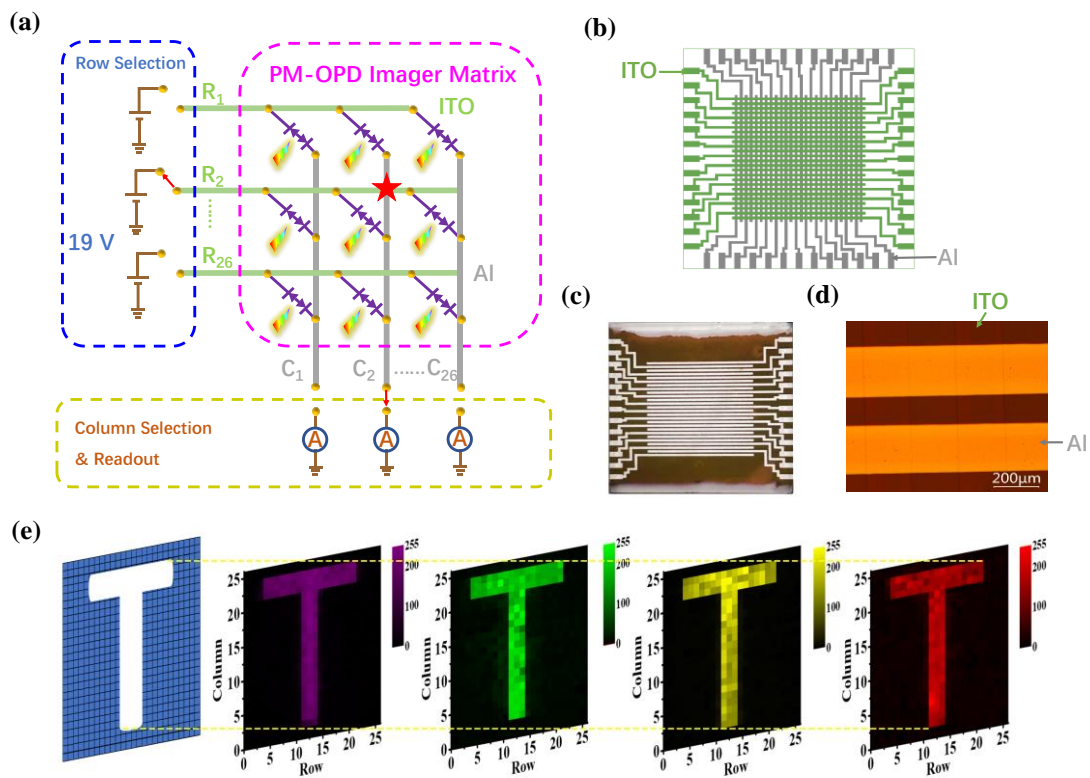

Fig. 5 (a) Schematic diagram for the pixel scanning process of the 26×26 image sensor based on the MIS based PM-OPD. (b) Schematic diagram of the crossed electrodes lines of the image sensor. (c-d) Physical map of the top cathode, and optical image of a part in the image sensor. (e) Imaging results of a "T" object under 375 nm, 505 nm, 565 nm, and 660 nm LED illumination with a power density of 1.27 mW/cm$^2$.

Table 1. Comparison of the performances of representative broad PM-OPDs made of polymer BHJs.

| Year [Ref] | Active layer | EQE@ Wavelength, bias | $R$ (A/W) | $D^*$ (Jones) | Response time | D: A weight ratio |
|---|---|---|---|---|---|---|
| 2017[36] | P3HT:PC$_{61}$BM | 2.08×10$^2$@550 nm, -0.5 V | 0.92 | 9.10×10$^{12}$ | - | |
| 2018[37] | PTAA:ZnO | 2.95×10$^2$@350 nm, -1 V | 83.20 | 3.70×10$^{12}$ | 110 ms | equal to 1 |
| 2019[38] | P3HT:PC$_{61}$BM:C$_{60}$ | 3.28×10$^2$@460 nm, -1 V | 1.21 | 4.22×10$^{12}$ | - | |
| 2020[39] | BT-2TPEPH:C$_{70}$ | 6.08×10$^4$@350 nm, -9 V | 1.72×10$^2$ | 3.08×10$^{12}$ | 11.69 ms | |
| 2012[40] | P3HT:ZnO | 3.41×10$^5$@360 nm, -9 V | 1.00×10$^3$ | 2.50×10$^{14}$ | 725 μs | |
| 2016[16] | PDPP3T:PC$_{71}$BM | 1.40×10$^5$@680 nm, 0.5 V | 7.62×10$^2$ | 6.30×10$^{12}$ | 270 ms | |
| 2018[41] | F8T2:ZnO NPs | 7.82×10$^2$@358 nm, 3 V | 2.24 | 8.45×10$^{12}$ | 12 ms | less than 1 |
| 2020[42] | PBDB-T:Y6 | 8.60×10$^3$@350 nm, 2 V | 24.27 | 2.13×10$^{12}$ | - | |
| 2020[43] | P3HT:PC$_{71}$BM | 3.90×10$^3$@355 nm, 5 V | 11.17 | 3.20×10$^{12}$ | - | |
| 2015[20] | P3HT:PC$_{71}$BM | 1.16×10$^5$@610 nm, -19 V | 5.69×10$^2$ | 2.17×10$^{14}$ | - | |
| 2016[17] | P3HT:ITIC | 1.17×10$^3$@625 nm, 15 V | 8.80 | 5.95×10$^{12}$ | - | |
| 2016[19] | P3HT:DC-IDT2T | 2.80×10$^4$@390 nm, -20 V | 1.31×10$^2$ | 1.43×10$^{14}$ | - | |
| 2019[18] | P3HT:PCBM | 1.20×10$^4$@450 nm, -15 V | 42 | 1.50×10$^{14}$ | > 1.14 s | larger than 1 |
| 2020[13] | P3HT:PCBM and P3HT:PTB7-Th:PCBM | 2.50×10$^3$@376 nm, 15 V | 7.47 | 8.00×10$^{12}$ | - | |
| 2020[15] | P3HT:ETBI-F | 1.56×10$^5$@350 nm, -20 V | 4.40×10$^2$ | 9.50×10$^{12}$ | 6.30 ms | |
| This work | P3HT:PC$_{71}$BM | 3.27×10$^6$@315 nm, 19 V | 8.29×10$^3$ | 6.76×10$^{14}$ | 83.59 ms | |

# Supporting Information

High-responsivity, High-detectivity Photomultiplication Organic Photodetector

Realized by a Metal-Insulator-Semiconductor Tunneling Junction

*Linlin Shi[1], Ning Li[2], Ting Ji[1], Ye Zhang[1], Wenyan Wang[1], Lin Feng[1], Rong Wen[1], Yuying Hao[1], Kaiying Wang[1,3], Furong Zhu[2], Guohui Li[1,\*], Beng S. Ong [2,\*], and Yanxia Cui[1,\*]*

[1]College of Physics and Optoelectronics, Key Lab of Interface Science and Engineering in Advanced Materials, Key Lab of Advanced Transducers and Intelligent Control System of Ministry of Education, Taiyuan University of Technology, Taiyuan 030024, China

[2]Department of Physics, Research Centre of Excellence for Organic Electronics, Institute of Advanced Materials, Hong Kong Baptist University, Kowloon Tong, Hong Kong, SAR, China

[3]Department of Microsystems, University of South-Eastern Norway, Horten, 3184, Norway

E-mail: yanxiacui@gmail.com, liguohui@tyut.edu.cn, bong@hkbu.edu.hk

**Experimental section**

The indium tin oxide (ITO)-coated glass substrates were sequentially cleaned in solution of detergent, deionized water, acetone, and isopropanol for 15 minutes in each round. Then, the cleaned ITO substrates were dried by air and subjected to the surface plasma treatment for 5 min to increase the work function of ITO surface. The PEDOT:PSS solution (purchased from Shanghai Han Feng Chemical Co., Ltd, Clevios™ PVP AI 4083) was spin-coated on the cleaned ITO substrate at 5000 rpm for 30 s, and then annealed at 120 °C for 15 min in the air. In the following, the PEDOT:PSS coated ITO substrates were transferred to the atomic layer deposition (ALD) integrated glove box to prepare the $Al_2O_3$ layer with different thicknesses of 0.2 nm, 0.4 nm, 0.6 nm, 0.8 nm, and 1 nm. The $Al_2O_3$ thin films were deposited at 150 °C using alternating exposures of $Al(CH_3)_3$ and $H_2O$ vapor in an ALD system. Polymethyl methacrylate (PMMA) was dissolved in ethyl acetate solvent to prepare 0.6 mg/mL solution, and then spin-coated on the surface of PEDOT:PSS or $Al_2O_3$. P3HT (Organtec Material Inc., the molecular weight and regularity are 5000 g/mol and 91%, respectively), PTB7 and $PC_{71}BM$ (1-Material, Corporation) were respectively dissolved in o-dichlorobenzene (o-DCB) solvent to prepare 40 mg/mL solution. The P3HT:$PC_{71}BM$ (PTB7:PCBM) solution was then prepared by mixing the P3HT (PTB7) and $PC_{71}BM$ solutions at a weight ratio of 100:1 and spin-coated onto the $Al_2O_3$ (PEDOT:SS or PMMA) layer at 1200 rpm for 30 s in the glove box (Jiangsu Lebo Science Instrument Co., Ltd.). The wet active layers were quickly annealed at 80 °C for 20 s, resulting in a 270 nm thick active layer. Finally, 100 nm Al cathode was thermally evaporated on the active layers in a high vacuum ($10^{-4}$ Pa) chamber (Shenyang Kecheng Vacuum Technology Co., Ltd.). The active area of each device is about 0.04 $cm^2$, which is defined by the overlap of the Al cathode and ITO anode. The preparation processes of 676 pixel for $Al_2O_3$ modified PM-OPD with active area of 0.04 $mm^2$ is the same as $Al_2O_3$ modified PM-OPD with the active area of 0.04 $cm^2$.

Dark and light current density versus voltage ($J$-$V$) characteristic curves of the PM-OPDs were measured by a programmable source meter (Keithley 2400). The EQE spectra of the PM-OPDs were measured under monochromatic light from a xenon lamp (Zolix, Gloria-X150A) coupled with a monochromator (Zolix, 0mni-λ 3005), with spectral power density in the μW/$cm^2$ level over the investigated wavelength range. The linear dynamic range (LDR) were measured by a programmable source meter (Keithley 2400) under a 532 nm continuous laser (Cnilaser, MGL-F-532) at different light intensities. The imaging was measured by a semiconductor analyzer (Agilent, B1500) under different wavelength LED (Thorlabs) with an intensity of 1.27 mW/$cm^2$. Electrochemical impedance spectroscopy (EIS) of the PM-OPDs were measured by an electrochemical workstation (Jiangsu Donghua Analytical Instruments Co., Ltd., DH700). Transient responses of the detectors were tested by a semiconductor analyzer (Agilent, B1500) with 532 nm laser (Cnilaser, MGL-F-532) controlled by a signal generator to produce pulsed light signals. Light intensities were measured by an optical power meter (Thorlabs, PM100A). The thicknesses of different functional layers were measured by a profiler (Bruker, Dektak XT). The optical image of the organic image sensor was characterized using an optical microscope (Nikon, LV-150).

**Calculations**

The external quantum efficiency and responsivity values of the PM-OPDs can be calculated according to the number of collected charges and the number of incident photons, as expressed by Equation (1-2) [1]:

$$EQE = \frac{I_{ph}/e}{P_{in}/h\nu} \quad (1)$$

$$R = \frac{I_{ph}}{P_{in}} = \frac{EQE \cdot e}{h\nu} \quad (2),$$

$I_{ph}$ is the photogenerated current (the measured current under light $(I_L)$ minus the dark current $(I_d)$), $e$ is the absolute value of the charge of the electron, $h$ is the Planck constant, and $\nu$ is the frequency of incident light, $h\nu$ is photon energy, and $P_{in}$ is the incident light power intensity.

The shot noise and thermal noise of PM-OPD can be calculated using the following equation:

$$i_{shot} = \sqrt{2eI_{dark}B} \quad (3)$$

$$i_{thermal} = \sqrt{\frac{4kTB}{R_{sh}}} \quad (4),$$

where $B$ is the bandwidth, $k$ is Boltzmann constant, $T$ is absolute temperature, and $R_{sh}$ is shunt resistance. If the total noise current dominated by the shot noise, the detectivity of the PM-OPDs can be calculated by shot noise as an approximation according to the following equation [2, 3].

$$D^* = \frac{R}{\sqrt{2eJ_d}} \quad (5)$$

The linear dynamic range of the devices are calculated by the following Equation:

$$LDR = 20\log\frac{P_{max}}{P_{min}} \quad (6),$$

where $P_{max}$ and $P_{min}$ are the maximum and minimum incident light intensities of the photocurrent density versus light intensity curve which lie within the linear response range.

The response time of the PM-OPDs reflect the response speed of the detector to receive incident light radiation, which is generally defined as the time for the photocurrent to rise from 10% to 90% of the highest value (fall from 90% to 10% of the highest value) during the on and off cycles of light illumination. The sum of the rise time and the falling time is counted as the response time of the photodetector.

In the sensing section, in order to reconstruct the optical pattern projected on the PM-OPD image matrix, we converted the photocurrent values to grayscale between 0 and 255 using the following conversion formula:

$$G = \frac{I_{meas}}{I_\Delta} \times 255 = \frac{I_{light} - I_{min}}{I_{max} - I_{min}} \times 255 \qquad (7),$$

where $I_{meas}$ is the measured photocurrent, which is the difference between the light current of each pixel and the minimum light current among all pixels. The $I_\Delta$ is the difference between the maximum and minimum light current among all pixels under different wavelength LED illumination.

**Simulations**

The theoretical method, two-dimensional (2D) Finite Element Method (FEM) was used to investigate the optical properties of the PM-OPD in this paper. All simulations were carried out assuming a periodic boundary along the z axis, which is normal to the film plane. Perfectly matched layer (PML) boundaries were applied at two planes perpendicular to the *z* axis, one lied in the glass and the other was in the air region next to the Al electrode. Light was illuminated from the ITO glass side. The wavelength dependent refractive indices and extinction coefficients of the P3HT:PC$_{70}$BM (100:1) layer were taken by the ellipsometer (Wuhan Eoptics Technology Co., Ltd, SE-VM). The refractive indices and extinction coefficients of other materials used in this work are extracted from literatures [4-6]. The distribution of absorption from the incident photons has a big impact on the performances of PM-OPDs, so we calculated the surface integration to obtain the wavelength-dependent absorption efficiency at different area of the active layer, and the corresponding absorption maps over the wavelength range between 300 nm and 800 nm was also plotted.

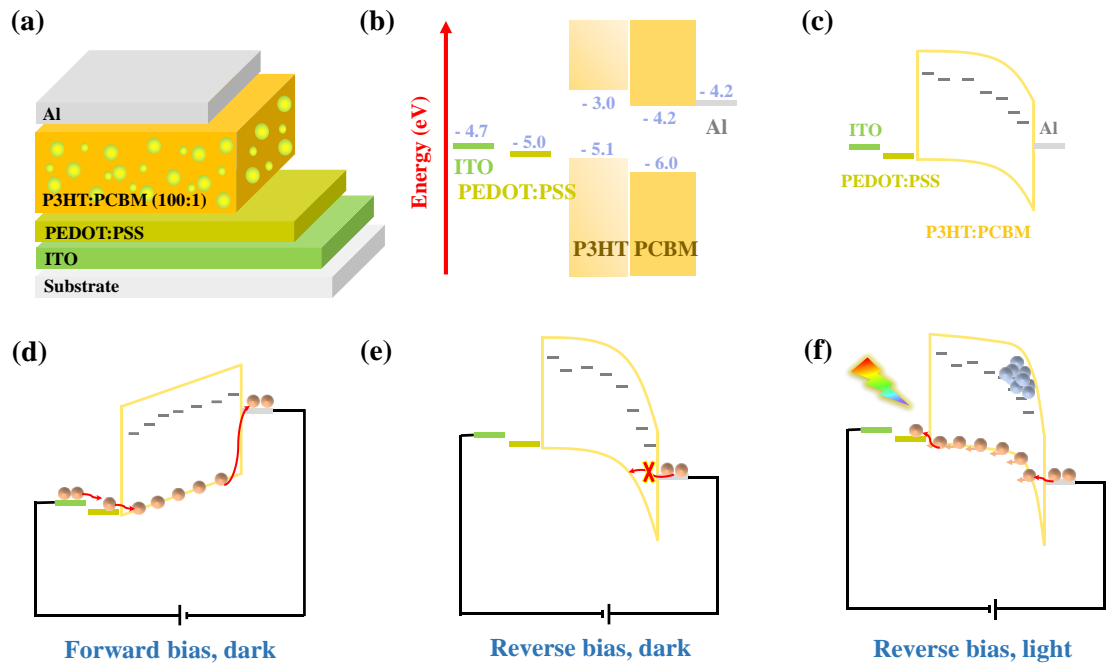

Figure S1 (a) Schematic diagram and (b) energy levels of used materials of the control PM-OPD; Energy band diagrams of the control PM-OPD when unbiased (c), forward biased (d) and reverse biased (e-f). The control PM-OPD is under dark and illumination in the diagrams of (d, e) and (f), respectively.

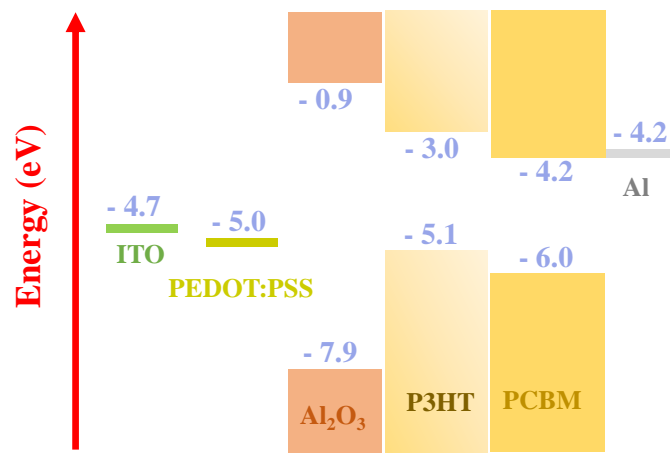

Figure S2 Energy levels of used materials in MIS tunneling junction based PM-OPD with $Al_2O_3$ layer.

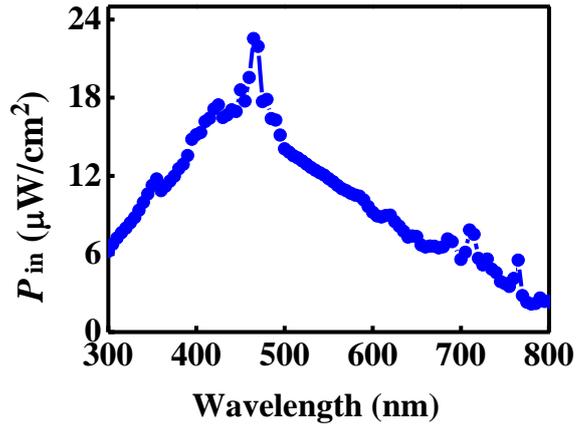

Figure S3 Incident light intensity spectrum of the monochromatic light used in external quantum efficiency measurements.

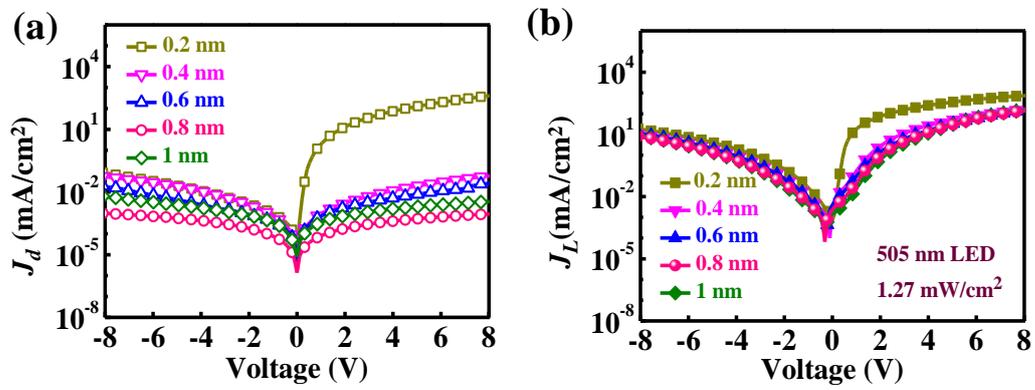

Figure S4 (a) Dark and (b) light $J$-$V$ curves under 505 nm at 1.27 mW/cm$^2$ illumination of the MIS tunneling junction based PM-OPD with different thickness of Al$_2$O$_3$ layer.

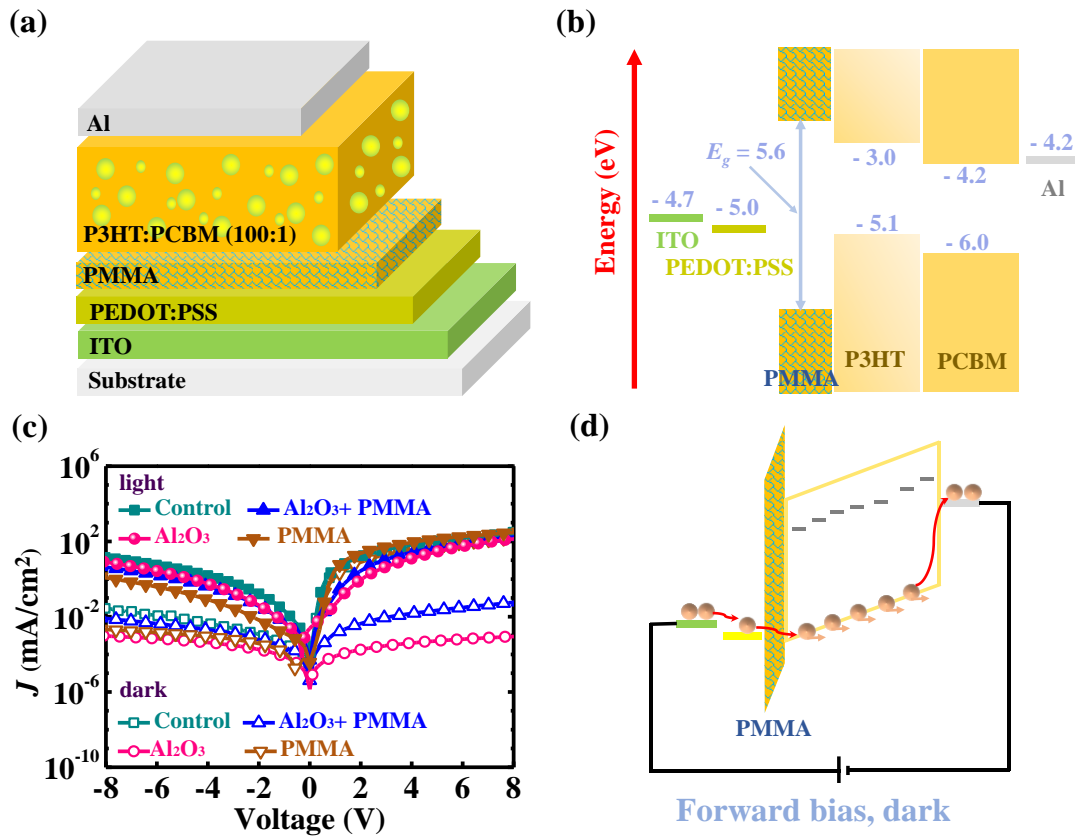

Figure S5 (a) Schematic diagram and (b) energy levels of used materials in the PM-OPD with PMMA interfacial layer; (c) The $J$–$V$ curves of the control PM-OPD and different interfacial layers modified PM-OPDs in dark and under 505 nm LED at 1.27 mW/cm$^2$; (d) Energy band diagrams of the PMMA modified PM-OPD in the dark and under forward bias.

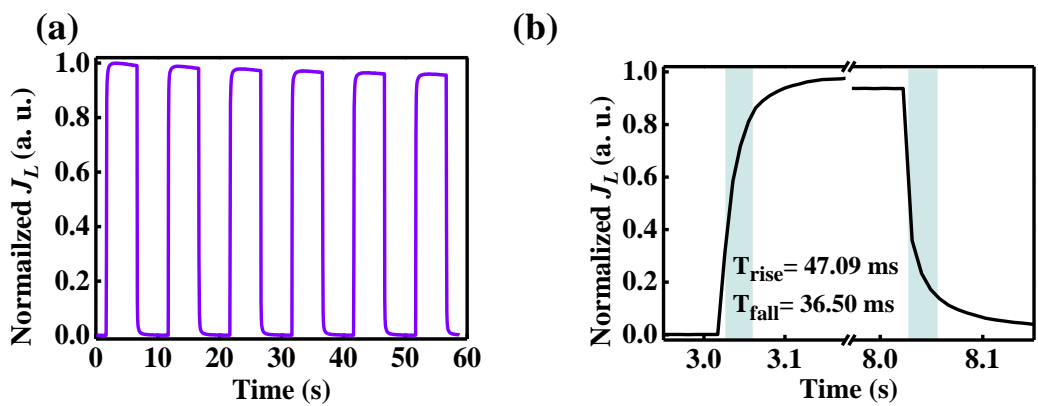

Figure S6 Normalized transient response of the MIS based PM-OPDs under illumination of 1.27 mW/cm$^2$.

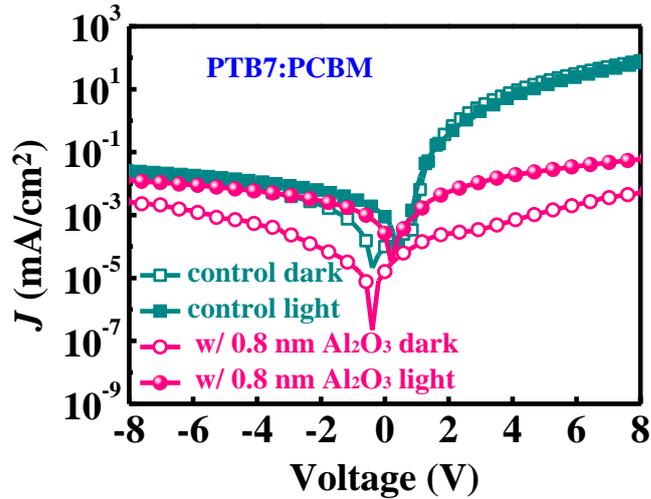

Figure S7 The *J–V* curves of the donor-rich PTB7:PCBM blend control OPD without interfacial layer and 0.8 nm thick $Al_2O_3$ modified OPD in dark and under 505 nm LED at 1.27 mW/cm$^2$.